\documentclass[12pt,a4paper]{article}
\usepackage[margin=25mm]{geometry}
\usepackage[T1]{fontenc}
\usepackage{lmodern,microtype}
\usepackage{amsmath,amssymb,amsthm,tikz}
\usepackage[authoryear,round,longnamesfirst]{natbib}

\usepackage{hyperref}

\hypersetup{pdftitle={Weak Atomic Continuity and Subgame Perfection},pdfauthor={Enxian Chen, Wei He, Yeneng Sun, and Hanping Xu}}
\newtheorem{theorem}{Theorem}
\newtheorem{lemma}{Lemma}
\newtheorem{proposition}{Proposition}
\newcommand{\In}{\mathrm{In}}
\newcommand{\Out}{\mathrm{Out}}
\newcommand{\contree}{%
\begin{tikzpicture}[x=1cm,y=1cm,font=\small,
  decision/.style={circle,fill=black,inner sep=1.4pt},
  terminal/.style={inner sep=0pt},
  action/.style={fill=white,inner sep=1.2pt},
  line width=0.45pt]
\node[decision,label=right:{$1$}] (row) at (0.0,0.00) {};
\node[decision,label=above left:{$2$}] (left) at (-3.2,-1.00) {};
\node[decision,label=above right:{$2$}] (right) at (3.2,-1.00) {};
\draw (row) -- node[action,above] {$L$} (left);
\draw (row) -- node[action,above] {$R$} (right);
\coordinate (ll) at (-4.8,-2.00);
\coordinate (lr) at (-1.6,-2.00);
\coordinate (rl) at (1.6,-2.00);
\coordinate (rr) at (4.8,-2.00);
\draw (left) -- node[action,above left] {$l$} (ll);
\draw (left) -- node[action,above right] {$r$} (lr);
\draw (right) -- node[action,above left] {$l$} (rl);
\draw (right) -- node[action,above right] {$r$} (rr);
\node[anchor=north,inner sep=3pt] at (ll) {$(-wz-d,\,wz)$};
\node[anchor=north,inner sep=3pt] at (lr) {$(2w-d,\,0)$};
\node[anchor=north,inner sep=3pt] at (rl) {$(wz-d,\,2w-wz)$};
\node[anchor=north,inner sep=3pt] at (rr) {$(-d,\,2w)$};
\end{tikzpicture}}

\title{Weak Atomic Continuity and Subgame Perfection}
\author{Enxian Chen\thanks{School of Economics, Nankai University, Tianjin, China 300071. E-mail: chenenxian777@gmail.com.}\qquad
Wei He\thanks{Department of Economics, The Chinese University of Hong Kong, Shatin, N.T., Hong Kong. E-mail: hewei@cuhk.edu.hk.}\qquad
Yeneng Sun\thanks{Departments of Economics and Mathematics, National University of Singapore, 10 Lower Kent Ridge Road, Singapore 119076. E-mail: ynsun@nus.edu.sg.}\qquad
Hanping Xu\thanks{Institute for Advanced Study in Mathematics, Harbin Institute of Technology, Harbin, China 150001. E-mail: e0321212@u.nus.edu.}}
\date{\today}
\begin{document}

\hypersetup{pageanchor=false}
\pagenumbering{gobble} 

\maketitle

\begin{abstract}
Subgame-perfect equilibrium need not exist in continuous games with perfect information when Nature moves. Under standard regularity assumptions, we prove that weak atomic continuity of Nature's transition probabilities yields a measurable pure-strategy subgame-perfect equilibrium for finite and infinite horizons. The condition, originating in the population genetics literature, is necessary for the existence of subgame-perfect equilibrium uniformly over the remaining primitives: whenever this condition is violated, one can construct a dynamic game with no subgame-perfect equilibrium, even in behavioral strategies. The proof preserves compactness and upper hemicontinuity of continuation payoff correspondences under integration without requiring convexity.


\end{abstract}

\newpage

\pagenumbering{arabic} 
\setcounter{page}{1}
\hypersetup{pageanchor=true}


\section{Introduction}\label{sec:introduction}

Dynamic games with perfect information are a basic framework for studying strategic decisions made in sequence. At each decision, the player who moves observes all previous actions and realizations of uncertainty. The study of such games dates to the analysis of chess by \citet{Zermelo1913}. Chess and Go are familiar examples; economic applications include bargaining with complete information, sequential investment, intergenerational transfers, and consumption with changing preferences. In each case, current decisions depend on anticipated future choices. The subgame perfection criterion requires these choices to be optimal after every history, including those that may never be reached.

The existence of subgame-perfect equilibrium is well understood in deterministic games. By the standard backward induction argument,  a pure-strategy subgame-perfect equilibrium exists in games with finite actions and finite horizons. \citet{Harris1985}, \citet{HL1987}, and \citet{HLRR1990}, among others, extend the equilibrium existence result to infinite-horizon continuous games under appropriate compactness and continuity assumptions. These extensions accommodate economic applications, where prices, quantities, and transfers are naturally continuous choices and interaction may continue indefinitely.

In dynamic settings, uncertainty is central to various economic applications, but it introduces new difficulties that change the existence problem. \citet{HRR1995} study perfect-information dynamic games with uncertainty. They provide an example to demonstrate that pure-strategy subgame-perfect equilibrium may fail to exist when Nature moves; they also propose an existence result in behavioral strategies. \citet{LM2003} show, however, that a continuous, four-player, five-stage game with perfect information and moves by Nature, in which the first two players choose actions from the unit interval, may fail to admit a subgame-perfect equilibrium, even in behavioral strategies. This raises a basic question: what conditions guarantee subgame-perfect equilibrium in dynamic games with perfect information?

To address the question above, we identify a key condition on Nature’s transitions, called \emph{weak atomic continuity}, that yields a general existence theorem. In particular, along a convergent sequence of histories, the condition requires that every atom of the limiting distribution be matched by atoms whose locations and probability weights converge to those of the limiting atom. The underlying topology was introduced by \citet{EK1994}, which is used to study convergence in population genetics while keeping track of the frequencies of individual genetic types.\footnote{In those models, an atom represents a genetic type with positive population frequency.} To our knowledge, it is the first application of this novel weak atomic topology  to an economic problem.

Under the condition of weak atomic continuity, we prove the existence of a  pure-strategy subgame-perfect equilibrium in both finite and infinite horizons. The payoff set of pure-strategy equilibrium is compact and varies upper hemicontinuously with the history. Nature’s state spaces may be noncompact, and intertemporal payoffs need not be stationary or additively separable. Our result establishes the perfect-information equilibrium existence in behavioral strategies proposed by \citet{HRR1995} under the appropriate condition on Nature’s transitions, and strengthens that conclusion to the existence of pure-strategy subgame-perfect equilibrium.

We further establish the necessity of weak atomic continuity for a general existence theorem. For any weakly continuous transition probability on a compact metric parameter space that violates the condition, we construct a two-player, five-stage game with no subgame-perfect equilibrium, even in behavioral strategies. The construction preserves the parameter space, state space, and transition probability. Every strategic choice after the initial move is binary, and every proper subgame has a pure-strategy subgame-perfect equilibrium. Whatever equilibrium play is given in these subgames, the first player has no optimal initial choice. Together, the two results provide a complete characterization under which a pure-strategy subgame-perfect equilibrium exists.

The economic difficulty arises because a realization of uncertainty can affect payoffs and provide a public event on which subsequent choices depend. Two states may become arbitrarily close while remaining distinct events. Players can choose different continuations after these events until the states merge at a limiting history. Atomless distributions can similarly shrink to a single state. Continuity of the payoff functions does not prevent this difficulty, because equilibrium continuation payoffs need not be continuous in the state. An expected continuation payoff attainable at nearby histories may become unattainable at the limit, leaving an earlier player with improving choices but no optimal choice. We provide an  example in Section~\ref{sec:example} to illustrate this issue.

Weak atomic continuity prevents this loss at limiting atoms while allowing both discrete and continuous uncertainty. Under the continuity assumptions, this condition is satisfied by deterministic transition probabilities, by transition probabilities on a fixed finite state space, and by those that are atomless after every history.

The existence proof rests on the following methodological observation. At a stage when Nature moves, expected payoffs are obtained by selecting an equilibrium continuation after each realized state and integrating the resulting payoffs against the transition probability. We prove that weak atomic continuity preserves nonemptiness, compactness, and upper hemicontinuity in this step. With only one active player at each strategic stage, these properties allow an optimal action to be supported by suitable equilibrium continuations, without requiring the convexity of the payoff sets.\footnote{\citet{HS2020} establish the existence of pure-strategy subgame-perfect equilibria in continuous games with perfect information under atomless transition probabilities. Atomlessness provides the required regularity and also makes the sets of expected continuation payoffs convex. Weak atomic continuity preserves the regularity needed for this step while allowing nonconvexities.}

The remainder of the paper is organized as follows. Section~\ref{sec:example} presents the illustrative example. Section~\ref{sec:model} describes the model, and Section~\ref{sec:results} introduces the continuity condition and states the results. Section~\ref{sec:conclusion} concludes. Proofs are collected in the Appendix.


\section{An illustrative example}\label{sec:example}

Consider the two-player, five-stage game $\Gamma^\epsilon$ in Figure~\ref{fig:example}, where $\epsilon\in[0,\tfrac12]$. Let
\[
D=\{0\}\cup\{1/m:m\geq1\},
\qquad S_\epsilon=(D+\epsilon)\cup(-D-\epsilon).
\]
Both spaces carry their inherited Euclidean topologies. Player~1 first chooses $x\in D$. Player~2 observes $x$ and chooses $\Out$ or $\In$. Choosing $\Out$ gives payoffs $(0,\tfrac12)$. After $\In$, Nature draws a publicly observed state $s$ according to
\[
K_x^\epsilon=\tfrac12\delta_{x+\epsilon}
                 +\tfrac12\delta_{-x-\epsilon}.
\]
Player~1 then chooses $L$ or $R$, and player~2 observes this choice before choosing $l$ or $r$. The initial choice $x$ enters player~1's payoff as a cost if player~2 enters. It also determines the distance between Nature's two realizations.

Nature's feasible set is $S_\epsilon$ after every $x$, including states assigned probability zero. After $\Out$, Nature draws from the same distribution and the players make the same binary choices, but payoffs remain $(0,\tfrac12)$. All action and state spaces are countable and compact, and terminal payoffs are bounded and continuous.

\begin{center}
\begin{minipage}{\linewidth}
\centering
\begin{tikzpicture}[x=1.5cm,y=1.1cm,line width=0.45pt,
 every node/.style={font=\small},
 dot/.style={circle,fill=black,inner sep=1.5pt}]
 \coordinate (root) at (-1.45,0);
 \coordinate (entry) at (-1.45,-0.78);
 \coordinate (out) at (-3.35,-1.55);
 \coordinate (nature) at (-0.10,-1.55);
 \coordinate (row) at (-0.10,-2.90);
 \coordinate (left) at (-2.60,-3.67);
 \coordinate (right) at (2.40,-3.67);
 \coordinate (ll) at (-3.75,-4.48);
 \coordinate (lr) at (-1.45,-4.48);
 \coordinate (rl) at (1.25,-4.48);
 \coordinate (rr) at (3.55,-4.48);
 \draw[densely dotted] (root) -- (-2.42,-0.43);
 \draw[densely dotted] (root) -- (-0.48,-0.43);
 \node at (-2.60,-0.43) {$\cdots$};
 \node at (-0.30,-0.43) {$\cdots$};
 \draw (root) -- node[right] {$x$} (entry);
 \draw (entry) -- node[above left] {$\Out$} (out);
 \draw (entry) -- node[above right] {$\In$} (nature);
 \draw[densely dotted] (nature) -- (-1.08,-1.98);
 \draw[densely dotted] (nature) -- (0.88,-1.98);
 \node at (-1.25,-1.98) {$\cdots$};
 \node at (1.05,-1.98) {$\cdots$};
 \draw (nature) -- node[pos=0.73,right] {$s\in\{x+\epsilon,-x-\epsilon\}$} (row);
 \draw (row) -- node[above left] {$L$} (left);
 \draw (row) -- node[above right] {$R$} (right);
 \draw (left) -- node[above left] {$l$} (ll);
 \draw (left) -- node[above right] {$r$} (lr);
 \draw (right) -- node[above left] {$l$} (rl);
 \draw (right) -- node[above right] {$r$} (rr);
 \node[dot,label=above:{$1$}] at (root) {};
 \node[dot,label=left:{$2$}] at (entry) {};
 \node[dot,label=right:{Nature, $K_x^\epsilon=\tfrac12\delta_{x+\epsilon}+\tfrac12\delta_{-x-\epsilon}$}] at (nature) {};
 \node[dot,label=right:{$1$}] at (row) {};
 \node[dot,label=above left:{$2$}] at (left) {};
 \node[dot,label=above right:{$2$}] at (right) {};
 \node[below left=-1pt] at (out) {$(0,\tfrac12)$};
 \node[below] at (ll) {$(-s-x,s)$};
 \node[below] at (lr) {$(2-x,0)$};
 \node[below] at (rl) {$(s-x,2-s)$};
 \node[below] at (rr) {$(-x,2)$};
\end{tikzpicture}
\makeatletter
\def\@captype{figure}
\makeatother
\caption{The five-stage game $\Gamma^\epsilon$.}
\label{fig:example}
\end{minipage}
\end{center}

Consider first $\epsilon\in(0,\tfrac12]$. After every initial choice, Nature draws a positive or a negative state, each with probability $\tfrac12$. At any $s>0$, player~2 strictly prefers $l$ after $L$ and $r$ after $R$. Player~1 compares $-s-x$ with $-x$ and strictly chooses $R$, giving payoffs $(-x,2)$. At any $s<0$, player~2 strictly prefers $r$ after $L$ and $l$ after $R$. Player~1 now compares $2-x$ with $s-x$ and strictly chooses $L$, giving payoffs $(2-x,0)$.

The expected payoff from $\In$ is
\[
\tfrac12(-x,2)+\tfrac12(2-x,0)=(1-x,1).
\]
Player~2 strictly prefers $\In$ to the payoff of $\tfrac12$ from $\Out$. Player~1 then maximizes $1-x$ by choosing $x=0$. Backward induction gives a pure-strategy subgame-perfect equilibrium with outcome
\[
x=0,\quad\In,\qquad
\begin{cases}
s=\epsilon,\ (R,r),&\text{with probability }\tfrac12,\\
s=-\epsilon,\ (L,r),&\text{with probability }\tfrac12.
\end{cases}
\]
At the initial choice $x=0$, Nature still produces two distinct public states, giving payoffs $(0,2)$ and $(2,0)$, with expectation $(1,1)$. The same sign-based choices apply at all other feasible states, including those of probability zero. Arbitrary pure choices complete the strategies after $\Out$. The strategies are Borel measurable because all history spaces are countable, and the backward-induction comparisons establish optimality against behavioral deviations at every history.

Now set $\epsilon=0$. At every $x>0$, Nature still draws a positive and a negative state. The same strict comparisons imply entry and payoffs $(1-x,1)$. At $x=0$, however, Nature draws only $s=0$. There is no additional public label distinguishing two latent draws of zero. Continuation choices must respond to this single observed state.

Player~2 is now indifferent between $l$ and $r$ after either $L$ or $R$. Let $q$ be the probability of $r$ after $L$ in any subgame-perfect continuation. Player~1 receives $2q$ from $L$ and zero from $R$. If $q>0$, player~1 strictly chooses $L$, leaving player~2 with zero from $\In$. Player~2 strictly prefers $\Out$, and player~1 receives zero. If $q=0$, player~1 receives zero regardless of the entry decision. Under every profile that is subgame perfect in all proper subgames, player~1's payoff from the initial choice must therefore be
\[
V^0(x)=
\begin{cases}
1-x,&x>0,\\
0,&x=0.
\end{cases}
\]
Its supremum is one, approached as $x=1/m\to0$, but no choice attains it. Every probability distribution over initial choices also gives an expected payoff strictly below one. A sufficiently large $m$ yields a profitable deviation to $x=1/m$, followed by the same continuation strategy. The game has no subgame-perfect equilibrium, even in behavioral strategies, although every proper subgame admits a pure-strategy subgame-perfect equilibrium.

Player~1 can make the cost $x$ arbitrarily small while preserving the two public events supporting entry. At zero cost those events merge. Any continuation that gives player~1 a positive payoff then leaves player~2 unwilling to enter. A positive $\epsilon$ keeps the events distinct even at $x=0$, allowing the initial player to attain the limiting payoff. The difficulty arises when distinct realizations of uncertainty become a single observed state and the continuation possibilities they supported disappear.


\section{Model}\label{sec:model}

There are finitely many players $I=\{1,\ldots,n\}$ and Nature, denoted by $0$. Let $H_0$ be a nonempty Polish space of initial histories and $X_{t+1}$ a Polish space of stage-$(t+1)$ moves. After a history $h\in H_t$, a designated mover $\iota_t\in I\cup\{0\}$ chooses $x\in D_t(h)\subseteq X_{t+1}$, giving
\[
H_{t+1}=\{(h,x):h\in H_t,\ x\in D_t(h)\}.
\]
The mover sequence $(\iota_t)$ is fixed. At each stage, exactly one player or Nature moves, and the move is observed before the next stage. There are no simultaneous moves or additional public randomization devices.

Each $D_t$ is upper and lower hemicontinuous with nonempty closed values, compact when $\iota_t\ne0$.\footnote{As a result, every $H_t$ is  Polish.} For a Polish space $S$, let $\mathcal P(S)$ denote its Borel probability distributions, endowed with the weak topology. At Nature's dates, a weakly continuous transition probability $K_t\colon H_t\to\mathcal P(X_{t+1})$ satisfies $K_t(h,D_t(h))=1$. Nature's feasible sets and the supports of its distributions need not be compact.

For horizon $T$, the payoff $u\colon H_T\to\mathbb R^n$ is bounded and continuous. For an infinite horizon, $u\colon Z\to\mathbb R^n$ is bounded and continuous, where
\[
Z=\{\omega=(h_0,x_1,x_2,\ldots):\omega|_t\in H_t\text{ for every }t\}
\]
has the product topology and $\omega|_t$ denotes the history through stage $t$. The closed feasibility graphs make $Z$ a closed, hence Polish, subspace of the ambient product. It need not be compact. Fix $C=[-M,M]^n$, which contains all payoffs. Exit branches may be padded by payoff-irrelevant moves.

 A correspondence $F\colon B\rightrightarrows Y$, where $B$ is a Borel subset of a Polish space and $Y$ is Polish, is measurable if $\{b:F(b)\cap O\ne\varnothing\}$ is Borel for every open $O\subseteq Y$.\footnote{Throughout the analysis, measurability  means Borel measurability.} A measurable selection is a Borel map $g$ satisfying $g(b)\in F(b)$ at every $b$.

A behavioral strategy for player $i$ consists of measurable transition probabilities $\beta_t\colon H_t\to\mathcal P(X_{t+1})$, supported on $D_t(h)$, at dates with $\iota_t=i$. A pure strategy consists of measurable selections from $D_t$, identified with Dirac measures. A profile $\beta$ is a subgame-perfect equilibrium (SPE) if, at every feasible history $h$,
\[
U_i(h;\beta)\ge U_i(h;\tau_i,\beta_{-i})
\]
for every player $i$ and measurable behavioral deviation $\tau_i$, where $U_i$ is expected payoff. This includes histories of probability zero. Let $E_t(h)$ be the set of pure-strategy SPE payoffs in the subgame at $h\in H_t$.


\section{Results}\label{sec:results}

This section identifies our key condition on Nature's transition probabilities. Under weak atomic continuity, a  pure-strategy subgame-perfect equilibrium exists for dynamic games in both finite and infinite horizons. The converse establishes that the condition is necessary for an existence result uniform over the remaining primitives. We begin by explaining the condition and its connection to the illustrative example.

\paragraph{Weak atomic continuity.}
Let $S$ be a Polish space. For $\mu\in\mathcal P(S)$, define
\begin{equation}\label{eq:atomic}
\mu^\star=\sum_{s:\mu\{s\}>0}\mu\{s\}^2\delta_s,
\qquad c(\mu)=\mu^\star(S),
\end{equation}
where $\delta_s$ is unit mass at $s$. The number $c(\mu)$ is the probability that two independent draws from $\mu$ coincide. A sequence $(\mu_m)$ converges \emph{weakly atomically} to $\mu$, written $\mu_m\Rightarrow_{\!a}\mu$, if
\[
\begin{gathered}
\int_S f\,d\mu_m\longrightarrow\int_S f\,d\mu
\quad\text{for every bounded continuous }f:S\to\mathbb R,\\
c(\mu_m)\longrightarrow c(\mu).
\end{gathered}
\]
A transition probability $K:H\to\mathcal P(S)$ on a Polish space $H$ satisfies \emph{weak atomic continuity} if $h_m\to h$ implies $K(h_m)\Rightarrow_{\!a}K(h)$. This is continuity in the weak atomic topology of \citet{EK1994}.\footnote{For a recent treatment of this topology, see \citet[Section~4.3]{DelloSchiavo2022}.}

The first convergence requirement alone defines ordinary \emph{weak convergence}, denoted by $\Rightarrow$. The weak atomic topology is finer than the weak topology, so weak atomic continuity implies weak continuity. Under weak convergence, convergence of $c(\mu_m)$ to $c(\mu)$ is equivalent to $\mu_m^\star\Rightarrow\mu^\star$ \citep[Lemmas~2.1--2.2]{EK1994}. It is also equivalent to the following matching property: every atom of the limiting probability distribution is approached by atoms whose locations and masses converge to those of that atom \citep[Lemma~2.5(a)]{EK1994}.

If $K(h)$ is atomless at every history, including the limiting history, then $c(K(h))=0$ throughout. For these transition probabilities, weak continuity and weak atomic continuity coincide. They also coincide for deterministic transition probabilities and for transition probabilities on a fixed finite state space. The difference between the two conditions arises when probability concentrates at a limiting atom without a matching atom along the approximating histories. If two states with probabilities $p,q>0$ merge, for example, their contribution to $c$ rises from $p^2+q^2$ to $(p+q)^2$.

Return to Section~\ref{sec:example}. For each fixed $\epsilon\in(0,\tfrac12]$, the two atoms of $K_x^\epsilon$ remain distinct as $x\to0$, and $c(K_x^\epsilon)=\tfrac12$ for every $x$. The transition probability satisfies weak atomic continuity. At $\epsilon=0$, $K_{1/m}^0\Rightarrow K_0^0=\delta_0$, but $c(K_{1/m}^0)=\tfrac12$ while $c(K_0^0)=1$. This transition probability is weakly continuous but does not satisfy weak atomic continuity. The merger removes the public distinction supporting entry, and the initial player's limiting payoff is no longer attainable.

\begin{theorem}\label{thm:existence}
If Nature's transition probabilities satisfy weak atomic continuity, a  pure-strategy SPE exists. At every date, $E_t$ has nonempty compact values and is upper hemicontinuous.
\end{theorem}


When a player moves, the single active player can optimize over compact sets of continuation payoffs without convexifying them. When Nature moves, these sets must be integrated over the next state. The following proposition shows that weak atomic continuity preserves the regularity properties needed for this step even when the continuation-payoff correspondence is nonconvex.

To state the integration result, let $H$ and $S$ be Polish, let $D\subseteq H\times S$ be closed, and let $K:H\to\mathcal P(S)$ be a transition probability satisfying $K(h,D(h))=1$, where $D(h)=\{s:(h,s)\in D\}$. For a correspondence $F:D\rightrightarrows C$, define
\begin{equation}\label{eq:integral}
\begin{aligned}
\mathcal I_KF(h)=\Bigl\{&\int_S g(s)K(h,ds):g:S\to C\text{ measurable},\\
&g(s)\in F(h,s)\text{ for }s\in D(h)\Bigr\}.
\end{aligned}
\end{equation}

\begin{proposition}\label{prop:integral}
If $K$ satisfies weak atomic continuity and $F$ has nonempty compact values and is upper hemicontinuous, then $\mathcal I_KF$ has these properties as well.
\end{proposition}

At a Nature date, take $H=H_t$, $S=X_{t+1}$, $D=H_{t+1}$, and $K=K_t$. If $F(h,s)$ is the set of continuation payoffs after state $s$, then $\mathcal I_{K_t}F(h)$ is the set of expected payoffs before Nature moves.

The converse concerns the existence result uniform over the remaining game primitives. Fix a compact metric space $H$, a Polish space $S$, and a weakly continuous transition probability $K:H\to\mathcal P(S)$. Consider games satisfying the assumptions of Section~\ref{sec:model} in which player~1 initially chooses $h\in H$ and Nature moves once, drawing a state in $S$ according to $K(h)$.

\begin{theorem}\label{thm:necessity}
If $K$ does not satisfy weak atomic continuity, there is a dynamic game  in which no SPE exists, even in behavioral strategies.
\end{theorem}

In the construction, we leave $H$, $S$, and $K$ unchanged, and construct a two-player, five-stage game. Every strategic choice after the initial move is binary, and every proper subgame has a pure-strategy SPE. The failure occurs at the initial choice, as in Section~\ref{sec:example}.

\begin{figure}[!ht]
\centering
\begin{tikzpicture}[x=1cm,y=1cm,font=\small,
  decision/.style={circle,fill=black,inner sep=1.4pt},
  terminal/.style={inner sep=0pt},
  action/.style={fill=white,inner sep=1.2pt},
  line width=0.45pt]
\node[decision,label=above:{$1$}] (root) at (0,0) {};
\node[decision,label=left:{$2$}] (entry) at (0,-1.0) {};
\draw (root) -- node[action,right] {$h$} (entry);
\draw[densely dotted] (root) -- (-1.25,-0.6);
\draw[densely dotted] (root) -- (1.25,-0.6);
\node at (-1.47,-0.6) {$\cdots$};
\node at (1.47,-0.6) {$\cdots$};
\node[terminal] (out) at (-2.7,-2.05) {$(0,c)$};
\node[decision,label=right:{Nature, $K(h)$}] (nature) at (1.7,-2.05) {};
\draw (entry) -- node[action,above left] {Out} (out);
\draw (entry) -- node[action,above right] {In} (nature);
\node[decision,label=right:{$1$}] (row) at (1.7,-3.15) {};
\draw (nature) -- node[action,left] {$s$} (row);
\draw[densely dotted] (nature) -- (0.45,-2.65);
\draw[densely dotted] (nature) -- (2.95,-2.65);
\node at (0.23,-2.65) {$\cdots$};
\node at (3.17,-2.65) {$\cdots$};
\node[decision,label=above left:{$2$}] (left) at (-1.5,-4.15) {};
\node[decision,label=above right:{$2$}] (right) at (4.9,-4.15) {};
\draw (row) -- node[action,above] {$L$} (left);
\draw (row) -- node[action,above] {$R$} (right);
\coordinate (ll) at (-3.1,-5.15);
\coordinate (lr) at (0.1,-5.15);
\coordinate (rl) at (3.3,-5.15);
\coordinate (rr) at (6.5,-5.15);
\draw (left) -- node[action,above left] {$l$} (ll);
\draw (left) -- node[action,above right] {$r$} (lr);
\draw (right) -- node[action,above left] {$l$} (rl);
\draw (right) -- node[action,above right] {$r$} (rr);
\node[anchor=north,inner sep=3pt] at (ll) {$(-wz-d,\,wz)$};
\node[anchor=north,inner sep=3pt] at (lr) {$(2w-d,\,0)$};
\node[anchor=north,inner sep=3pt] at (rl) {$(wz-d,\,2w-wz)$};
\node[anchor=north,inner sep=3pt] at (rr) {$(-d,\,2w)$};
\end{tikzpicture}
\caption{The five-stage game.}
\label{fig:necessity}
\end{figure}
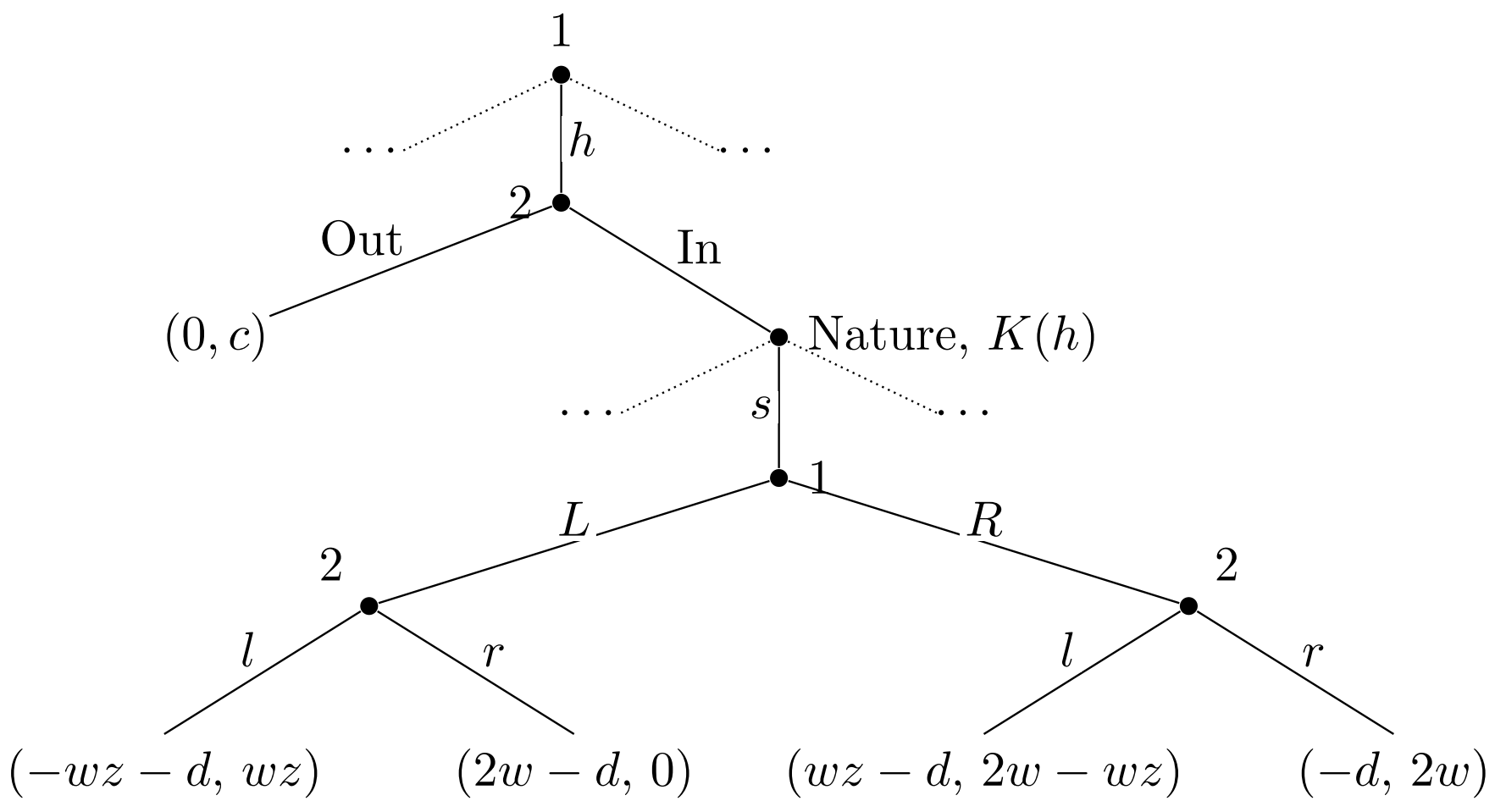

Figure~\ref{fig:necessity} uses the sequence of moves from Section~\ref{sec:example}. There is a single initial history. Player~1 chooses $h\in H$, and Nature draws the state according to the probability distribution $K(h)$ after either entry decision. Payoffs remain $(0,c)$ throughout the remaining stages after $\Out$.

The parameters are $c>0$ and continuous functions $w\colon S\to[0,1]$, $z\colon H\times S\to[-1,1]$, and $d\colon H\to[0,\infty)$. The weight $w$ concentrates the relevant payoffs near a limiting atom, $z$ determines the continuation incentives, and $d$ is player~1's cost upon entry. In the figure, $w=w(s)$, $z=z(h,s)$, and $d=d(h)$. The proof chooses these functions so that player~1's payoff approaches a supremum along a convergent sequence of initial choices but stays strictly below it at every choice.

\section{Conclusion}\label{sec:conclusion}

We establish a general existence theorem for dynamic games with perfect information under weak atomic continuity and prove its necessity for existence uniformly over the remaining game primitives. The condition ensures that limits of attainable continuation payoffs remain attainable as Nature's probability distributions vary with the history. 
Our integration result preserves compactness and upper hemicontinuity of the continuation-payoff correspondence without requiring convexity, allowing the construction of a pure-strategy subgame-perfect equilibrium. 

\appendix

\section{Appendix}\label{app:proofs}

We use the measurable selection and maximum theorems in \citet{AB2006}. For a compact-valued correspondence into the compact payoff set $C$, upper hemicontinuity is equivalent to a closed graph and implies measurability.

\subsection{Proof of Proposition~\ref{prop:integral}}

\begin{proof}
The measurable selection theorem ensures that $\mathcal I_KF(h)$ is nonempty at every $h$. Since all expected payoffs lie in the compact set $C$, it remains to show that $\mathcal I_KF$ has a closed graph. Let $h_m\to h$ and $r_m\to r$, with $r_m\in \mathcal I_K F(h_m)$. Choose measurable $g_m:S\to C$ with $g_m(s)\in F(h_m,s)$ on $D(h_m)$ and $\int g_m(s)K(h_m,ds)=r_m$. Put $\mu_m=K(h_m)$ and $\mu=K(h)$.

The joint distributions of states and selected payoffs,
\[
\gamma_m(ds,dq)=\mu_m(ds)\delta_{g_m(s)}(dq),
\]
are uniformly tight by weak convergence of $\mu_m$ and compactness of $C$. Pass to a subsequence with $\gamma_m\Rightarrow\gamma$. Testing bounded continuous functions $\varphi:S\to\mathbb R$ and the payoff coordinates gives
\[
\int\varphi(s)\gamma(ds,dq)=\lim_m\int\varphi\,d\mu_m=\int\varphi\,d\mu,
\qquad
\int q\,d\gamma=\lim_m\int q\,d\gamma_m=r.
\]
The limit $\gamma$ has state marginal $\mu$ and mean payoff $r$. The graph of $F$, viewed as a subset of $H\times S\times C$, is closed. Since $\delta_{h_m}\otimes\gamma_m\Rightarrow\delta_h\otimes\gamma$, Portmanteau gives $(\delta_h\otimes\gamma)(\operatorname{Gr}F)=1$.

We show that $\gamma$ assigns a single payoff to each atom of $\mu$. List these atoms as $s_j$, with masses $p_j$. By \citet[Lemma~2.5(a)]{EK1994}, there are atoms $s_{m,j}\to s_j$ with masses $p_{m,j}\to p_j$, defined eventually for each $j$. Every matched atom lies in $D(h_m)$ because $K(h_m,D(h_m))=1$. Compactness of $C$ permits successive subsequences on which $g_m(s_{m,j})$ converges for $j=1,2,\ldots$. A diagonal subsequence retains convergence for every fixed $j$. Closedness of the graph of $F$ then gives
\[
g_m(s_{m,j})\longrightarrow q_j\in F(h,s_j).
\]
For every bounded continuous $\psi\ge0$ on $S\times C$,
\[
\int\psi\,d\gamma=\lim_m\int\psi\,d\gamma_m
\ge\lim_m p_{m,j}\psi(s_{m,j},g_m(s_{m,j}))=p_j\psi(s_j,q_j).
\]
Take $\psi_k(y)=\max\{0,1-k\operatorname{dist}(y,(s_j,q_j))\}$ in a compatible metric and let $k\to\infty$. This yields $\gamma\{(s_j,q_j)\}\ge p_j$. Since the state marginal has mass $p_j$ at $s_j$, equality holds and all mass above $s_j$ is at $q_j$.

Let $\nu=\mu-\sum_jp_j\delta_{s_j}$ be the atomless part of $\mu$, and disintegrate $\gamma$ as $\mu(ds)\kappa_s(dq)$. The support property gives $\int q\,\kappa_s(dq)\in\operatorname{co} F(h,s)$ for $\mu$-almost every $s$, where $\operatorname{co}$ denotes convex hull. Extend $F(h,\cdot)$ by $\{0\}$ outside $D(h)$. If $\nu\ne0$, Aumann's integral theorem (\citealp{Aumann1965}), applied to the completed atomless measure space, gives a selection $g$ of $F(h,\cdot)$ with
\[
\int g\,d\nu=\iint q\,\kappa_s(dq)\,\nu(ds).
\]
Replace $g$ by a Borel version and, on the Borel null set where it is infeasible, by a measurable selection from $F(h,\cdot)$. If $\nu=0$, start with any such selection. Set $g(s_j)=q_j$ at the atoms and $g=0$ outside $D(h)$. These changes preserve its $\nu$-integral and make it Borel and feasible everywhere on $D(h)$. Hence
\[
\int g\,d\mu=\sum_j p_jq_j+\int g\,d\nu=\int q\,d\gamma=r.
\]
This proves $r\in \mathcal I_K F(h)$.
\end{proof}

\subsection{Feasible continuation distributions}

For the infinite horizon, let $\mathcal L_t(h)\subseteq\mathcal P(Z)$ be the set of probability distributions on $Z$ induced from history $h$ by feasible behavioral strategy profiles, with Nature's transition probabilities fixed. Here $\mathcal P(Z)$ carries the weak topology. Each distribution is supported on paths extending $h$.

\begin{lemma}\label{lem:feasible}
Under the assumptions of Section~\ref{sec:model}:
\begin{enumerate}
\renewcommand{\labelenumi}{(\roman{enumi})}
\item $\mathcal L_t$ is nonempty, compact-valued, and upper hemicontinuous. Its union over every compact subset of $H_t$ is compact.
\item Any measurable choice $P_\eta\in\mathcal L_N(\eta)$ can be implemented, at every $\eta\in H_N$, by a single measurable continuation profile, and pasted after any feasible distribution of earlier play.
\item There are bounded continuous $u_N:H_N\to C$ such that, for every $t$ and compact $B\subseteq H_t$,
\begin{equation}\label{eq:approx}
\sup_{h\in B}\sup_{P\in\mathcal L_t(h)}\int
\|u(\omega)-u_N(\omega|_N)\|_\infty\,P(d\omega)\longrightarrow0.
\end{equation}
\end{enumerate}
\end{lemma}

\begin{proof}
(i) Feasible profiles exist by measurable selection at strategic dates. Fix a compact set $B\subseteq H_t$. We first show that the distributions induced from $B$ are uniformly tight. Starting with $B$ at date $t$, construct compact sets of histories at successive dates. Given a compact set of histories at date $k$, the union of feasible strategic moves over that set is compact at a strategic date, by compact-valued upper hemicontinuity. At a Nature date, the distributions $K_k(h)$ over that set form a weakly compact family and are uniformly tight. Given $\varepsilon>0$, at each date $k\ge t$ we can choose a compact set of next moves such that, conditional on any history in the current compact set, the probability of leaving the chosen sets at that step is at most $\varepsilon 2^{-(k-t+1)}$. Intersect the resulting product of $B$ and the compact sets of moves with the closed feasible path space. This gives a compact set of paths with probability at least $1-\varepsilon$ under every feasible profile starting in $B$.

To show closedness, take $P_m\in\mathcal L_t(h_m)$ with $h_m\in B$ and $P_m\Rightarrow P$. Along a subsequence, $h_m\to h\in B$. The date-$t$ marginal of $P$ is $\delta_h$. For each Nature date $k\ge t$, each bounded continuous $f$ on $H_k$, and each bounded continuous $g$ on $X_{k+1}$, the identity
\[
\int_Z f(\omega|_k)
\left[g(x_{k+1})-\int_{X_{k+1}}g(x)K_k(\omega|_k,dx)\right]P_m(d\omega)=0
\]
passes to $P$ by weak continuity of $K_k$. Products of bounded continuous functions determine the joint distribution on $H_k\times X_{k+1}$. These identities identify the conditional distribution of $x_{k+1}$ given $\omega|_k$ as $K_k(\omega|_k)$, almost surely under $P$. At each strategic date, disintegrate the corresponding marginal of $P$ to obtain a Borel conditional distribution. It is feasible almost surely since $P$ is on $Z$. At histories where it is infeasible, substitute the Dirac measure at a measurable feasible move. The resulting profile induces $P$ from $h$ by successive conditioning. This proves closedness of $\bigcup_{h\in B}\mathcal L_t(h)$, and uniform tightness gives compactness. The same argument shows that if $h_m\to h$ and $P_m\in\mathcal L_t(h_m)$, then a subsequence converges to a member of $\mathcal L_t(h)$. This gives upper hemicontinuity and compact values.

(ii) At a strategic date $k\ge N$, consider the marginal of $P_\eta$ on the current history $h\in H_k$ and next move $x\in X_{k+1}$.  Lemma~12 in \citet{HRR1995} gives a conditional distribution $\beta_k(dx\mid\eta,h)$ that is jointly Borel in the starting history $\eta$ and current history $h$. Where $\beta_k(D_k(h)\mid\eta,h)<1$, replace it by a Dirac measure at a measurable feasible selection. For each $\eta$, this changes the conditional distribution only on a Borel set of histories having probability zero under $P_\eta$.

At history $h$, the continuation profile uses $\beta_k(\cdot\mid h|_N,h)$, since the starting history $\eta=h|_N$ is observed as part of $h$. The strategic conditional distributions then agree with those of every $P_\eta$. Nature's transition probabilities are fixed, so the profile reproduces each $P_\eta$. Applying this profile after any feasible earlier play proves pasting.

(iii) By the Tietze extension theorem, extend $u$ from the closed set $Z$ to a continuous function $\widetilde u:H_0\times\prod_{k\ge1}X_k\to C$. Choose reference points $\bar x_k\in X_k$ and define
\[
u_N(h_0,x_1,\ldots,x_N)=\widetilde u(h_0,x_1,\ldots,x_N,\bar x_{N+1},\bar x_{N+2},\ldots).
\]
This is a bounded continuous function on $H_N$. If $A\subseteq Z$ is compact, all these reference-tail paths and the original paths in $A$ lie in the compact product of $A$'s coordinate projections with the reference points added. Their distances from the original paths tend uniformly to zero in a product metric. Uniform continuity of $\widetilde u$ on that compact product therefore gives
\[
\sup_{\omega\in A}\|u(\omega)-u_N(\omega|_N)\|_\infty\longrightarrow0.
\]
For a compact $B\subseteq H_t$, part (i) supplies, for every $\varepsilon>0$, a compact $A\subseteq Z$ with $P(A)>1-\varepsilon$ for every $P\in\bigcup_{h\in B}\mathcal L_t(h)$. For every such $P$,
\[
\int\|u(\omega)-u_N(\omega|_N)\|_\infty\,P(d\omega)
\le\sup_{\omega\in A}\|u(\omega)-u_N(\omega|_N)\|_\infty+2M\varepsilon.
\]
Letting $N\to\infty$ and then $\varepsilon\downarrow0$ proves \eqref{eq:approx}.
\end{proof}

\subsection{Proof of Theorem~\ref{thm:existence}}

\begin{proof}
\emph{Finite horizon. Backward induction.} Let $F:H_{t+1}\rightrightarrows C$ be nonempty, compact-valued, and upper hemicontinuous. At a strategic date with mover $i$, put
\[
\theta_i(h)=\sup_{x\in D_t(h)}\min_{q\in F(h,x)}q_i.
\]
Define the backward operator at every date by
\begin{equation}\label{eq:backward}
O_tF(h)=
\begin{cases}
\{r:\exists a\in D_t(h),\ r\in F(h,a),\ r_i\ge\theta_i(h)\}, & \iota_t=i\ne0,\\
\mathcal I_{K_t}F(h), & \iota_t=0.
\end{cases}
\end{equation}
At a strategic date, action $a$ and payoff $r$ can be supported by assigning an $i$-minimizing continuation after every other action. The inequality $r_i\ge\theta_i(h)$ rules out a profitable deviation.

The minimum over $F(h,x)$ is lower semicontinuous in $(h,x)$. Lower hemicontinuity of $D_t$ therefore makes $\theta_i$ lower semicontinuous. Hence the supporting triples $(h,a,r)$ in \eqref{eq:backward} form a closed set. If $(h_m,r_m)\to(h,r)$ with $r_m\in O_tF(h_m)$, compactness and upper hemicontinuity of $D_t$ give a convergent subsequence of supporting actions. The limiting action supports $r$ at history $h$, so the correspondence $O_tF$ has a closed graph. Nonemptiness follows by maximizing $q_i$ over the compact graph of $F(h,\cdot)$ on $D_t(h)$. The maximizing payoff is at least $\theta_i(h)$. At Nature's dates, Proposition~\ref{prop:integral} applies. The operator $O_t$ preserves nonemptiness, compactness, and upper hemicontinuity. It is also monotone, since enlarging $F$ enlarges its integral and lowers the strategic threshold.

For horizon $T$, define
\begin{equation}\label{eq:finite}
Q_T(h)=\{u(h)\},\qquad Q_t=O_tQ_{t+1}\quad(t<T).
\end{equation}

\emph{Forward induction.} Take a measurable selection $v_t\in O_tF$. At a strategic date, choose measurably a supporting action $\sigma_t(h)$ in \eqref{eq:backward}. Set $v_{t+1}(h,\sigma_t(h))=v_t(h)$, and select an $i$-minimizing payoff from $F(h,x)$ elsewhere. Both correspondences have nonempty compact values and Borel graphs, and hence admit Borel selections. Since $\sigma_t$ has a Borel graph, $v_{t+1}$ is measurable.

At a Nature date, extend $F(h,x)$ by $\{0\}$ outside $H_{t+1}$. The extension is measurable, with nonempty compact values in $C$. Apply \citet[Section~2, part~(3)]{Mertens2003} with parameter $h$, state $x$, and transition probability $K_t$. His theorem provides a jointly Borel selection $f(h,r,x)$, defined for $r\in\mathcal I_{K_t}F(h)$, such that
\[
f(h,r,x)\in F(h,x),\qquad \int f(h,r,x)K_t(h,dx)=r
\]
for every feasible $(h,x)$ and every such $r$. Consequently, $v_{t+1}(h,x)=f(h,v_t(h),x)$ is jointly Borel and satisfies
\begin{equation}\label{eq:selection}
v_{t+1}(h,x)\in F(h,x),\qquad \int v_{t+1}(h,x)K_t(h,dx)=v_t(h)
\end{equation}
for every $h\in H_t$ and $x\in D_t(h)$.

Starting with a selection $v_t$ of $Q_t$, apply these constructions at every subsequent history to obtain a pure profile $\sigma$ and selections $v_k$ of $Q_k$. At a strategic history $h$ with mover $i$, the prescribed action gives $v_k(h)$, while every other move $x$ gives, using the threshold for $F=Q_{k+1}$,
\[
v_{k+1,i}(h,x)=\min_{q\in Q_{k+1}(h,x)}q_i\le\theta_i(h)\le v_{k,i}(h).
\]
The same inequality holds after averaging over any behavioral choice of $x$. At another player's date the prescribed action gives equality for all coordinates, and at Nature's dates \eqref{eq:selection} gives equality in conditional expectation. Hence under $\sigma$ the conditional expectation of $v_{k+1}$ is $v_k$, and under any unilateral behavioral deviation by $i$ its $i$th coordinate is at most $v_{k,i}$. Since $v_T=u$, iterated expectation proves that $\sigma$ realizes $v_t$ and is an SPE at every history.

Any $r\in Q_t(h)$ is realizable by changing the starting selection at $h$ to $r$. Conversely, backward induction places every pure-strategy SPE payoff in $Q_t$. At a strategic date optimality implies \eqref{eq:backward}, and at a Nature date the payoff is an integral of continuation payoffs. It follows that $E_t=Q_t$.

\emph{Infinite horizon. Payoff correspondences.} For each date, define the coordinate bounds on feasible continuation payoffs:
\begin{equation}\label{eq:box}
\begin{gathered}
R_t(h)=\prod_{i\in I}[\ell_{ti}(h),b_{ti}(h)],\\
\ell_{ti}(h)=\min_{P\in\mathcal L_t(h)}\int u_i\,dP,\qquad
b_{ti}(h)=\max_{P\in\mathcal L_t(h)}\int u_i\,dP.
\end{gathered}
\end{equation}
The box contains every feasible payoff, although its coordinates need not be jointly attainable. The bounds $\ell_{ti}$ and $b_{ti}$ are lower and upper semicontinuous, respectively, by Lemma~\ref{lem:feasible}(i). The graph of $R_t$ is therefore closed. Indeed, if $h_m\to h$ and $q_m\to q$ with $q_m\in R_t(h_m)$, lower semicontinuity of $\ell_{ti}$ gives $\ell_{ti}(h)\le q_i$, and upper semicontinuity of $b_{ti}$ gives $q_i\le b_{ti}(h)$. Since $R_t$ has nonempty closed values in the compact set $C$, it is compact-valued and upper hemicontinuous.

We have $O_tR_{t+1}\subseteq R_t$. At a strategic date, any distribution from $(h,x)$ extends to one from $h$ by choosing $x$, so $R_{t+1}(h,x)\subseteq R_t(h)$. At a Nature date, choose measurably distributions attaining each coordinate's lower and upper bounds at successor histories. Pasting them by Lemma~\ref{lem:feasible}(ii) gives
\[
\ell_{ti}(h)\le\int\ell_{t+1,i}(h,x)K_t(h,dx)
\le\int b_{t+1,i}(h,x)K_t(h,dx)\le b_{ti}(h).
\]
Every integral of a selection from $R_{t+1}$ lies in $R_t$.

For each truncation date $N$, put $Q_N^N=R_N$ and $Q_t^N=O_tQ_{t+1}^N$ for $t<N$. Since $O_NR_{N+1}\subseteq R_N$, monotonicity gives $Q_t^{N+1}\subseteq Q_t^N\subseteq R_t$. It follows that
\[
Q_t=\bigcap_{N\ge t}Q_t^N
\]
is nonempty, compact-valued, and upper hemicontinuous. We show that $Q_t=O_tQ_{t+1}$.

Fix $h$ and, for $N\ge t+1$, write $F_N(x)=Q_{t+1}^N(h,x)$ and $F_\infty(x)=\bigcap_{N\ge t+1}F_N(x)$. At a Nature date, apply Proposition~\ref{prop:integral} with parameter space $\{0\}\cup\{1/N:N>t\}$, state space $D_t(h)$, and constant distribution $K_t(h)$. Use $F_N$ at $1/N$ and $F_\infty$ at zero. This correspondence has a closed graph. Along a sequence approaching parameter zero, every selected payoff eventually belongs to each fixed $F_k$, so its limit belongs to every $F_k$. If $r$ belongs to all the integrals on the left below, use the constant sequence of expected payoffs $r$ at parameters $1/N$. Proposition~\ref{prop:integral} places its limit in the integral at zero. The reverse inclusion follows from $F_\infty\subseteq F_N$. Hence
\[
\bigcap_{N>t} \mathcal I_{K_t}Q_{t+1}^N(h)=\mathcal I_{K_t}Q_{t+1}(h).
\]

At a strategic date with mover $i$, the minima over $F_N(x)$ increase to the minimum over $F_\infty(x)$ for each fixed $x$. A convergent subsequence of minimizers exists in the compact set $R_{t+1}(h,x)$, and nesting puts its limit in every $F_N(x)$. Interchanging the two suprema gives
\[
\sup_{N>t}\sup_{x\in D_t(h)}\min_{q\in F_N(x)}q_i
=\sup_{x\in D_t(h)}\sup_{N>t}\min_{q\in F_N(x)}q_i
=\sup_{x\in D_t(h)}\min_{q\in F_\infty(x)}q_i.
\]
For $r\in Q_t(h)$, its coordinate $r_i$ is at least each threshold on the left, and hence the limiting threshold on the right. Take a convergent subsequence of its supporting actions $a_N\in D_t(h)$. For every fixed $k$, eventually $r\in F_k(a_N)$. Closedness puts $r\in F_k(a)$ at the limit action $a$. It follows that $r\in F_\infty(a)$ and $r\in O_tQ_{t+1}(h)$. The reverse inclusion follows from monotonicity, proving the recursion.

\emph{Verification.} Forward induction gives selections $v_t\in Q_t$ and a pure profile $\sigma$ with the conditional equalities and deviation inequalities already established. To pass from $v_N$ to $u$, use $u_N$ from Lemma~\ref{lem:feasible}(iii) and define
\[
e_N(\eta)=\max_{P\in\mathcal L_N(\eta)}\int\|u(\omega)-u_N(\eta)\|_\infty\,P(d\omega),
\qquad \eta\in H_N.
\]
The objective is bounded and continuous, so Lemma~\ref{lem:feasible}(i) and measurable maximization give a Borel choice of maximizing distribution. For each coordinate $i$ and feasible distribution $P$ from $\eta$,
\[
\left|\int u_i\,dP-u_{N,i}(\eta)\right|\le e_N(\eta).
\]
Both endpoints of the $i$th interval in \eqref{eq:box} therefore lie within $e_N(\eta)$ of $u_{N,i}(\eta)$. It follows that $\|q-u_N(\eta)\|_\infty\le e_N(\eta)$ for every $q\in R_N(\eta)$.

Fix $h\in H_t$, $N\ge t$, and $P\in\mathcal L_t(h)$. Keep its distribution through date $N$ and paste the maximizing distributions thereafter. The resulting $P'\in\mathcal L_t(h)$ has the same date-$N$ marginal, so
\[
\int e_N(\omega|_N)\,dP=\int\|u(\omega)-u_N(\omega|_N)\|_\infty\,dP'.
\]
Since $v_N\in Q_N\subseteq R_N$, the triangle inequality and \eqref{eq:approx} imply
\begin{equation}\label{eq:payofflimit}
\begin{aligned}
\sup_{P\in\mathcal L_t(h)}\int\|v_N(\omega|_N)-u(\omega)\|_\infty\,dP
&\le 2\sup_{P\in\mathcal L_t(h)}\int\|u(\omega)-u_N(\omega|_N)\|_\infty\,dP\\
&\longrightarrow0.
\end{aligned}
\end{equation}
For every unilateral behavioral deviation $\tau_i$, iterated conditional expectation gives
\[
\mathbb E_h^{\tau_i,\sigma_{-i}}[v_{N,i}(\omega|_N)]
\le v_{t,i}(h)=\mathbb E_h^\sigma[v_{N,i}(\omega|_N)].
\]
Both induced distributions lie in $\mathcal L_t(h)$, so \eqref{eq:payofflimit} permits passage to $u$. The profile $\sigma$ therefore realizes $v_t$ and is an SPE in every subgame. Conversely, a pure-strategy SPE has date-$N$ continuation payoffs in $R_N$. Backward induction puts its date-$t$ payoff in every $Q_t^N$, hence in $Q_t$. Continuation payoffs are measurable by integration against the induced path distributions. As in the finite horizon, any point of $Q_t(h)$ can be realized. Therefore $E_t=Q_t$.
\end{proof}

\subsection{Proof of Theorem~\ref{thm:necessity}}

For a finite measure $\mu$ and a bounded nonnegative measurable function $w$, let $w\mu$ denote the measure defined by $(w\mu)(A)=\int_A w\,d\mu$ for every Borel set $A$.

\begin{lemma}\label{lem:split}
Let $S$ be Polish and let $(\mu_m)_{m\ge 1}$ be a sequence in $\mathcal P(S)$ with $\mu_m\Rightarrow\mu$ but $\mu_m^\star\not\Rightarrow\mu^\star$. After passing to a subsequence and indexing it again by $m=1,2,\ldots$, there exist an atom $s_0$ of $\mu$, a continuous function $w:S\to[0,1]$ with $w(s_0)=1$, a constant $c>0$, and continuous functions $g_m:S\to[-1,1]$ such that, for $\nu_m=w\mu_m$ and $\nu=w\mu$,
\begin{equation}\label{eq:split}
\begin{gathered}
\nu(S\setminus\{s_0\})<c/2,\qquad \nu_m\{g_m=0\}=0,\qquad \nu_m\{g_m>0\}\ge c,\\
\alpha_m:=\nu_m\{g_m<0\}\longrightarrow\alpha\ge c.
\end{gathered}
\end{equation}
\end{lemma}

\begin{proof}
Fix a compatible metric $d_S$ on $S$. Since $0\le\mu_m^\star\le\mu_m$ and $(\mu_m)$ is uniformly tight, the measures $\mu_m^\star$ are uniformly tight and have bounded total mass. Their failure to converge to $\mu^\star$ therefore gives a subsequence with $\mu_m^\star\Rightarrow\lambda\ne\mu^\star$. For every bounded nonnegative continuous $v$ on $S$, the function $(s,s')\mapsto v(s)\mathbf 1_{\{s=s'\}}$ is bounded and upper semicontinuous because the diagonal is closed. Since $\mu_m\otimes\mu_m\Rightarrow\mu\otimes\mu$, the Portmanteau inequality gives
\[
\int v\,d\lambda
=\lim_m\int v(s)\mathbf 1_{\{s=s'\}}\,d(\mu_m\otimes\mu_m)(s,s')
\le\int v\,d\mu^\star.
\]
For a nonempty closed set $F$, approximating its indicator by $\max\{0,1-kd_S(\cdot,F)\}$ gives $\lambda(F)\le\mu^\star(F)$. Inner regularity gives $\lambda\le\mu^\star$. Since the two measures differ and $\mu^\star$ is concentrated on the atoms of $\mu$, some $s_0$ satisfies $p:=\mu\{s_0\}>0$ and $\lambda\{s_0\}<p^2$.

Choose $\delta\in(0,p)$ with $\lambda\{s_0\}<(p-\delta)^2$. Take a ball $B=B_S(s_0,r)$ small enough that $\lambda(B)<(p-\delta)^2$ and $\mu(B\setminus\{s_0\})<\delta/16$, with $\lambda(\partial B)=0$. Such a radius exists because the balls shrink to $\{s_0\}$ and only countably many spheres have positive $\lambda$-mass. Set $w(s)=\max\{0,1-2d_S(s,s_0)/r\}$. Its support is contained in $B$, and $w(s_0)=1$. For all sufficiently large $m$, $\mu_m^\star(B)<(p-\delta)^2$, so every atom of $\mu_m$ in $B$ has mass below $p-\delta$. The same bound holds for every atom of $\nu_m$. Continuity of $w$ gives $\nu_m(S)\to\nu(S)\ge p$. Discard finitely many terms so that $\nu_m(S)\ge p-\delta/16$.

For each $m$, choose a Borel set $E_m$ with $\delta/4\le\nu_m(E_m)\le p-\delta/4$. If an atom has mass at least $\delta/4$, take that atom. Otherwise take mass $\delta/4$ from the atomless part if possible. If the atomless part has smaller mass, take all of it and add atoms until the total first reaches $\delta/4$. The total is then below $\delta/2$, since each atom has mass below $\delta/4$. By inner regularity choose disjoint compact sets $F_m\subseteq E_m$ and $G_m\subseteq S\setminus E_m$ such that $\nu_m(S\setminus(F_m\cup G_m))<\delta/16$. Both sets are nonempty by the bounds on $\nu_m(E_m)$ and $\nu_m(S)$.

Define
\[
g_m(s)=\frac{d_S(s,F_m)-d_S(s,G_m)}{2(d_S(s,F_m)+d_S(s,G_m))}-t_m,
\qquad t_m\in(-1/4,1/4).
\]
The denominator is positive because the sets are disjoint and closed. The function is continuous, lies in $[-1,1]$, is negative on $F_m$, and is positive on $G_m$. The zero sets for distinct $t_m$ are disjoint, so at most countably many choices have positive $\nu_m$-mass. Choose $t_m$ with $\nu_m\{g_m=0\}=0$. The sets $E_m$ and $\{g_m<0\}$ differ only outside $F_m\cup G_m$. It follows that $3\delta/16\le\alpha_m\le p-3\delta/16$, and $\nu_m\{g_m>0\}=\nu_m(S)-\alpha_m\ge\delta/8$. Set $c=\delta/8$ and pass to a further subsequence along which $\alpha_m\to\alpha\ge c$. Finally, $\nu(S\setminus\{s_0\})\le\mu(B\setminus\{s_0\})<\delta/16=c/2$.
\end{proof}

\begin{proof}[Proof of Theorem~\ref{thm:necessity}]
Since $K$ is weakly continuous and does not satisfy weak atomic continuity, there are distinct $h_m\ne h_0$ with $h_m\to h_0$ such that $K(h_m)^\star$ does not converge weakly to $K(h_0)^\star$. Apply Lemma~\ref{lem:split} to $\mu_m=K(h_m)$ and $\mu=K(h_0)$. Relabel its subsequence by $m=1,2,\ldots$. Take $s_0$, $c$, $w$, $(g_m)$, $(\alpha_m)$, and $\alpha$ from that lemma. For every $h\in H$, write $\nu_h=wK(h)$, so $\nu_{h_m}=\nu_m$ and $\nu_{h_0}=\nu$. We use $c$ as player~2's payoff from $\Out$ and $w$ as the state-dependent payoff weight in Figure~\ref{fig:necessity}. We choose $z$ and $d$ so that, for every profile that is an SPE in all proper subgames, player~1's payoff approaches $2\alpha$ along $(h_m)$ and is strictly below $2\alpha$ at every initial choice.

\medskip
Step 1. Fix a compatible metric $d_H\le1$ on $H$. By taking a further subsequence, assume $r_{m+1}<r_m/4$, where $r_m=d_H(h_m,h_0)>0$. Keep the corresponding $g_m$ and $\alpha_m$ and index them again by $m=1,2,\ldots$. For each $m$, $w\mathbf1_{\{g_m\le0\}}$ is bounded and upper semicontinuous. By weak continuity of $K$ and the Portmanteau inequality, $h\mapsto\nu_h\{g_m\le0\}$ is upper semicontinuous. At $h_m$ its value is $\alpha_m$, since Lemma~\ref{lem:split} gives $\nu_m\{g_m=0\}=0$. Choose an open neighborhood $V_m$ of $h_m$ such that $\overline V_m\subset B_H(h_m,r_m/8)$ and
\begin{equation}\label{eq:local}
\nu_h\{g_m\le0\}<\alpha_m+1/m
\qquad(h\in V_m).
\end{equation}
The balls are disjoint, since their distances from $h_0$ lie between $7r_m/8$ and $9r_m/8$ and $r_{m+1}<r_m/4$. They eventually lie in every neighborhood of $h_0$. For each $m$, choose a continuous $\phi_m\colon H\to[0,1]$ with $\phi_m(h_m)=1$ and support contained in $V_m$.

Set $d_m=4(\alpha_m-\alpha)_++6/m$, where $x_+=\max\{x,0\}$. Define
\begin{equation}\label{eq:construction}
\begin{aligned}
z(h,s)&=d_H(h,h_0)\left[1+\sum_{m=1}^{\infty}\phi_m(h)(g_m(s)-1)\right],\\
d(h)&=\sum_{m=1}^{\infty} d_m\phi_m(h).
\end{aligned}
\end{equation}
At most one summand is nonzero for each $h$. In $V_m$, the bracket equals $1-\phi_m(h)+\phi_m(h)g_m(s)$. Because $g_m(s)\ge-1$, it is strictly positive if $\phi_m(h)<1/2$. A negative or zero value therefore requires $\phi_m(h)\ge1/2$, in which case $d(h)\ge d_m/2$. The chosen cost satisfies
\begin{equation}\label{eq:cost}
2(\alpha_m+1/m)-d_m/2=2\min\{\alpha_m,\alpha\}-1/m<2\alpha.
\end{equation}

The sums in \eqref{eq:construction} are locally finite away from $h_0$, so both functions are continuous there. The bracket lies in $[-1,1]$, and hence $|z(h,s)|\le d_H(h,h_0)$ for every $s$. This gives joint continuity at every $(h_0,s)$, with $z(h_0,s)=0$. For $d$, fix $\varepsilon>0$ and choose $m_0$ such that $d_m<\varepsilon$ for $m>m_0$, which is possible because $\alpha_m\to\alpha$. A neighborhood of $h_0$ avoids the supports of $\phi_1,\ldots,\phi_{m_0}$. Throughout this neighborhood $0\le d(h)<\varepsilon$, since at most one remaining term can contribute. The function $d$ is continuous at $h_0$, with $d(h_0)=0$. At $h_m$, \eqref{eq:construction} gives $z(h_m,s)=r_mg_m(s)$ and $d(h_m)=d_m$. Since $r_m>0$, $z(h_m,s)$ and $g_m(s)$ have the same sign and the same zeros. The payoffs are bounded and continuous. The feasible sets are constant, with compact sets at each player's move, and Nature's distribution after the initial choice is the given $K(h)$.

\medskip
Step 2. Fix $h$ and suppose player~2 has chosen $\In$ and Nature has drawn $s$. The remaining part of Figure~\ref{fig:necessity} is reproduced below. Write $w=w(s)$, $z=z(h,s)$, and $d=d(h)$.
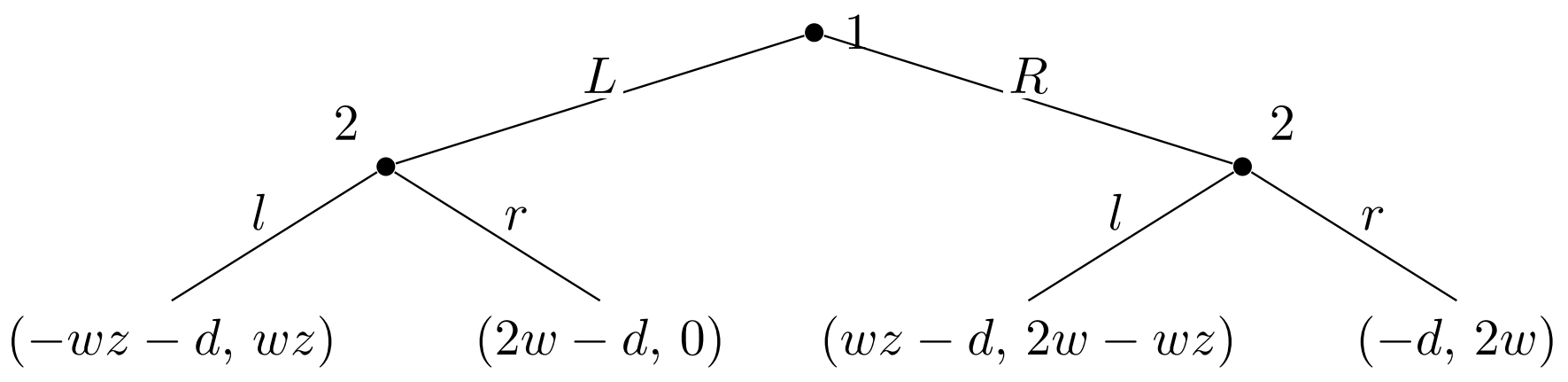
\begin{center}
\contree
\end{center}

If $w=0$, all four terminal payoff vectors are $(-d,0)$. Suppose $w>0$. If $z>0$, player~2 strictly prefers $l$ after $L$ and $r$ after $R$. Player~1 strictly prefers $R$, since $-d>-wz-d$. If $z<0$, player~2 strictly prefers $r$ after $L$ and $l$ after $R$. Player~1 strictly prefers $L$, since $2w-d>wz-d$.

If $z=0$, player~2 is indifferent at both of her decision nodes. Let $\beta$ be the probability that she chooses $r$ after $L$. Player~1 gets $2w\beta-d$ from $L$ and $-d$ from $R$. If $\beta>0$, player~1 strictly chooses $L$ and player~2 receives zero. If $\beta=0$, player~1 is indifferent between $L$ and $R$. Player~1 receives $-d$, and player~2 receives $2w$ times the probability of $R$.

Every proper subgame admits a pure-strategy SPE. The last two stages admit Borel pure choices by the preceding comparisons, with a fixed action at ties. Integrating player~2's resulting payoff over $K(h)$ gives a Borel function of $h$. At the entry decision, she can choose $\In$ if that function exceeds $c$ and $\Out$ otherwise. This yields Borel strategies that are an SPE in every proper subgame.

Now fix any Borel behavioral strategy profile that is an SPE in every proper subgame. Let $q(h,s)$ be the payoff vector for players~1 and~2 after $(h,\In,s)$. Let $J(h)=\int_S q(h,s)K(h,ds)$ be their expected payoff vector conditional on $(h,\In)$, and let $V(h)$ be player~1's payoff after $h$, before player~2 chooses whether to enter. Since $\Out$ gives player~1 zero,
\begin{equation}\label{eq:entry}
V(h)\le\max\{0,J_1(h)\}.
\end{equation}
The comparisons above give the following payoff possibilities, including the case $w=0$:
\begin{equation}\label{eq:cases}
q(h,s)\in
\begin{cases}
\{(-d,2w)\}, & z>0,\\
\{(2w-d,0)\}, & z<0,\\
([-d,2w-d]\times\{0\})\cup(\{-d\}\times[0,2w]), & z=0.
\end{cases}
\end{equation}
In each case, $q_1(h,s)\le 2w(s)\mathbf1_{\{z(h,s)\le0\}}-d(h)$. Since player~1 pays $d(h)$ upon entry regardless of Nature's draw, integrating gives
\begin{equation}\label{eq:upper}
J_1(h)\le 2\nu_h\{s:z(h,s)\le0\}-d(h).
\end{equation}
If $\nu_h\{s:z(h,s)=0\}=0$, the zero set contributes no gross payoff. In that case, \eqref{eq:cases} gives
\begin{equation}\label{eq:exact}
J_1(h)=2\nu_h\{s:z(h,s)<0\}-d(h),
\qquad J_2(h)=2\nu_h\{s:z(h,s)>0\}.
\end{equation}

At $h_m$, \eqref{eq:construction} gives $z(h_m,s)=r_mg_m(s)$ and $d(h_m)=d_m$. By Lemma~\ref{lem:split}, $\nu_m\{g_m=0\}=0$, $\nu_m\{g_m<0\}=\alpha_m$, and $\nu_m\{g_m>0\}\ge c$. Substituting these into \eqref{eq:exact} gives $J_1(h_m)=2\alpha_m-d_m$ and $J_2(h_m)\ge2c>c$. Player~2 strictly chooses $\In$, so
\begin{equation}\label{eq:sequence}
V(h_m)=2\alpha_m-d_m=2\alpha-2|\alpha_m-\alpha|-6/m\longrightarrow 2\alpha.
\end{equation}

At $h_0$, \eqref{eq:construction} gives $z(h_0,s)=0$ for every $s$ and $d(h_0)=0$. By \eqref{eq:cases}, each player's payoff at any state $s$ is bounded above by $2w(s)$. Hence states other than $s_0$ contribute at most $2\nu(S\setminus\{s_0\})<c$ to either player's expected payoff, by Lemma~\ref{lem:split}. If $J_1(h_0)>c$, this bound implies $q_1(h_0,s_0)>0$. The last case of \eqref{eq:cases} then gives $q_2(h_0,s_0)=0$. Player~2 therefore earns $J_2(h_0)<c$ from entry and strictly chooses $\Out$, giving player~1 zero. If $J_1(h_0)\le c$, \eqref{eq:entry} gives $V(h_0)\le c$. In either case, since $\alpha\ge c>0$ by Lemma~\ref{lem:split},
\begin{equation}\label{eq:limit}
V(h_0)\le c<2\alpha.
\end{equation}

\medskip
Step 3. Fix $h\ne h_0$. If $\phi_m(h)<1/2$ for every $m$, then $z(h,s)>0$ for every $s$. By \eqref{eq:cases}, $J_1(h)=-d(h)\le0$, so \eqref{eq:entry} gives $V(h)\le0<2\alpha$.

Otherwise there is a unique $m$ with $\phi_m(h)\ge1/2$. Then $h\in V_m$ and $d(h)\ge d_m/2$. The bracket in \eqref{eq:construction} is positive wherever $g_m(s)>0$, so $\{s:z(h,s)\le0\}\subseteq\{s:g_m(s)\le0\}$. By \eqref{eq:upper}, \eqref{eq:local}, and \eqref{eq:cost},
\[
J_1(h)\le 2\nu_h\{g_m\le0\}-d(h)
<2(\alpha_m+1/m)-d_m/2<2\alpha.
\]
Equation~\eqref{eq:entry} again gives $V(h)<2\alpha$. Together with \eqref{eq:limit}, this holds for every $h\in H$.

The function $V$ is bounded and Borel by iterated integration. For any distribution $\zeta$ over initial choices, player~1's expected payoff $\int_H V(h)\,\zeta(dh)$ is strictly below $2\alpha$, since $2\alpha-V(h)>0$ at every $h$. By \eqref{eq:sequence}, some $h_m$ gives a strictly higher payoff. Player~1 can choose that $h_m$ and follow the same continuation strategy. No initial strategy is optimal under any profile that is an SPE in all proper subgames, and the game has no behavioral SPE.
\end{proof}

\end{document}